\documentclass[acmsmall]{acmart}
\usepackage[utf8]{inputenc}
\AtBeginDocument{%
  }
\usepackage{fvextra}
\usepackage{pdftexcmds}
\usepackage{graphicx}
\graphicspath{ {./generated/} }

\usepackage{subcaption}
\usepackage{cleveref}
\usepackage{paralist}
\usepackage{listings}
\usepackage[most]{tcolorbox}
\usepackage{csquotes}
\usepackage{enumitem}  
\usepackage{relsize}
\usepackage{tikz}
\usepackage{wrapfig}
\usepackage{xspace}

\SetExpansion[context = sloppy,shrink = 10000]{encoding = {OT1,T1,TS1} }{}

\lstnewenvironment{notes}
    {\lstset{basicstyle=\rmfamily,columns=fullflexible,breaklines=true,escapechar=\$}}
    {}

\definecolor{main}{HTML}{b3b3b3}    
\definecolor{sub}{HTML}{f1f1f1}     

\tcbset{
    sharp corners,
    colback = white,
    before skip = 0.5em,    
    after skip = 1em      
}                           

\newtcolorbox{boxResult}{
    colback = sub, 
    colframe = main, 
    boxrule = 0pt, 
    toprule = 1pt, 
    bottomrule = 1pt 
}

\newcommand{\markerRaw}{\tikz[baseline]{%
\node[rectangle,rounded corners=0.4mm,draw=black!80,inner sep=2pt,anchor=base] (A) {Raw};}}

\newcommand{\markerCleaned}{\tikz[baseline]{%
\node[rectangle,rounded corners=0.4mm,draw=black!80,inner sep=2pt,anchor=base] (A) {Clean};}}

\newcommand{\markerFiltered}{\tikz[baseline]{%
\node[rectangle,rounded corners=0.4mm,draw=black!80,inner sep=2pt,anchor=base] (A) {Filter};}}

\DeclareRobustCommand\circled[1]{\tikz[baseline=(char.base)]{
    \node[shape=circle,draw,inner sep=1pt] (char) {#1};}}

\newcommand{\datasetDOI}{doi:10.5281/zenodo.14000867}

\newcommand{\repoURL}{https://github.com/CISPA-SysSec/fuzz-harness-degradation}

\def\CPP{{C\nolinebreak[4]\hspace{-.05em}\raisebox{.35ex}{\footnotesize\bf ++}}\xspace}

\setcopyright{cc}
\setcctype{by-nc-nd}
\acmDOI{10.1145/3808172}
\acmYear{2026}
\acmJournal{PACMSE}
\acmVolume{3}
\acmNumber{FSE}
\acmArticle{FSE165}
\acmMonth{7}
\acmSubmissionID{fse26mainb-p1142-p}
\received{2025-09-11}
\received[accepted]{2026-03-24}

\begin{document}

\include{generated/latex-data}

\title{An Empirical Study of Fuzz Harness Degradation}

\author{Philipp Görz}
\orcid{0009-0001-0501-1249}
\affiliation{%
  \institution{Ruhr-University Bochum}
  \city{Bochum}
  \country{Germany}
}
\email{philipp.goerz@rub.de}

\author{Joschua Schilling}
\orcid{0009-0000-2097-9891}
\affiliation{%
  \institution{CISPA Helmholtz Center for Information Security}
  \city{Saarbrücken}
  \country{Germany}
}
\email{joschua.schilling@cispa.de}

\author{Nicolai Bissantz}
\orcid{0000-0001-7301-4567}
\affiliation{%
  \institution{Ruhr-University Bochum}
  \city{Bochum}
  \country{Germany}
}
\email{nicolaibernhard.bissantz@rub.de}

\author{Thorsten Holz}
\orcid{0000-0002-2783-1264}
\affiliation{%
  \institution{MPI-SP}
  \city{Bochum}
  \country{Germany}
}
\email{thorsten.holz@mpi-sp.org}



\begin{abstract}

    Fuzzing is a widely used technique to automatically test software for potential faults. To fuzz software projects efficiently and effectively, software developers must use \emph{fuzz harnesses}, i.e., small programs that connect the fuzzer to the project's code under test.
    However, as projects evolve, it is unclear whether fuzz harnesses are maintained in lockstep or left to stagnate, and whether unmaintained fuzz harnesses gradually degrade in terms of code coverage and bug-finding effectiveness.

    In this paper, we focus on OSS-Fuzz, the largest continuous fuzzing platform in practice, which provides harnesses for 510 security-critical open-source C/\CPP projects. These harnesses are usually contributed by project maintainers or external developers, yet their ongoing maintenance is not always ensured.
    Our analysis shows that, overall, harnesses exhibit only a small reduction in coverage and retain surprising longevity in their ability to uncover bugs.
    At the same time, we also identify cases where harnesses degrade, analyze their root causes and the involved semantics of the code changes, and categorize them systematically. 
    Finally, we extend OSS-Fuzz and Fuzz Introspector, a companion project to investigate fuzzer performance, with new metrics to automatically detect harness degradation, enabling more effective monitoring of fuzzing quality in evolving projects.

\end{abstract}

\maketitle

\section{Introduction}

In recent years, fuzzing has become a popular technique in software security. By running programs on huge amounts of automatically generated inputs, fuzzers have uncovered an astonishing number of software faults and security vulnerabilities in widely used software.
For example, until May 2025, Google's continuous fuzzing platform OSS-Fuzz has helped identify and fix over 13,000 vulnerabilities and 50,000 other bugs across approximately 1,000 open-source projects~\cite{serebryany2017ossFuzzTalk, ossFuzzRepo}. 

To fuzz software efficiently, projects must be integrated via a \emph{fuzzing harness} (or fuzz harness for short), i.e., a small piece of glue code that feeds inputs generated by the fuzzer into the program under test in the correct manner. Creating a good harness requires considerable effort and knowledge of the target program~\cite{liang2018fuzzChallengesPractice, nourry2023human}.
Despite automation attempts~\cite{jung2021winnieHarnessSynthesis, jeon2023spHarnessSynthesis, babic2019fudgeHarnessSynthesis, ispoglou2020fuzzgenHarnessSynthesis, jeong2023utopia, ossFuzzGenJavaPaper, zhang2021apicraft, lyu2023promptFuzzingForDriverGen, zhang2021intelligen, zhang2023daisyFuzzDriverGenDynamic, zhang2023automataGuidedGen}, most harnesses are still manually written and maintained.

\begin{wrapfigure}{r}{0.4\textwidth}
    \centering
    \includegraphics[width=\linewidth]{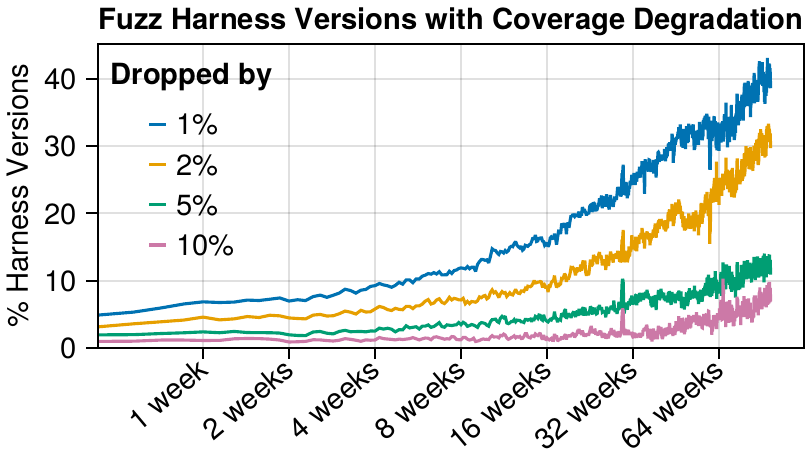}
    \caption{Share of C/\CPP projects in OSS-Fuzz projects with different levels of coverage degradation over time.}
    \label{fig:intro-plot}
    \vspace{-2ex}
\end{wrapfigure}

The importance of a well-designed harness cannot be overstated: An effective harness maximizes the fuzzer's reach in the code, while a poor harness may only yield superficial testing. With this in mind, Google even offers financial incentives for OSS-Fuzz integrations, rewarding projects that contribute fuzzing harnesses that cover at least 50\% of their code~\cite{ossFuzzRewardProgram}. This reward program underscores that higher coverage leads to more effective and longer-running fuzz campaigns, as fuzzers often stall when constrained by an incomplete harness. In fact, previous work has observed that a persistent plateau in coverage in a fuzzing campaign is often caused by an inadequate harness design that prohibits the fuzzer from reaching deeper into the target~\cite{gao2023beyondCoveragePlateauFuzzBlockers}.

However, one crucial question remains unanswered: \emph{What happens to a fuzzing harness after initial integration when the software being tested undergoes further development?} Modern software is highly dynamic: New features are added regularly, code is redesigned, and dependencies are changed~\cite{agileDev1, agileDev2, agileDev3}. Ideally, the fuzz harness should be updated in parallel with these changes to continue testing new code. In practice, however, the harness and the main code base are often developed independently, which can lead to the harness becoming obsolete. The target program may outgrow the original scope of the harness, or changes to APIs may render parts of the harness ineffective. This phenomenon can be viewed as a special case of software/test suite \emph{degradation}, in which components of a system gradually lose effectiveness as the system evolves. The effect of varying rates of evolution within an ecosystem of software projects is well studied under software degradation~\cite{softwareDegradation1, softwareDegradation2, softwareDegradation3, softwareDegradation4}. It is noteworthy that the fuzzing community has paid little attention to this problem so far, even though such degradation could unnoticeably reduce the fuzzer's ability to find security-critical bugs over time.

Silent harness degradation is particularly sneaky because it often goes unnoticed until a bug slips through. A fuzz target that continues to ``run'' continuously can give project maintainers a false sense of security, even if the actual code coverage has decreased. For example, a security review of the popular cURL project in 2024 revealed that the developers ``had accidentally drastically shrunk the fuzzing coverage a while back without even noticing''~\cite{curl-harness}. This example illustrates the general problem: a silent degradation of fuzz harnesses can undermine the very security benefits that continuous fuzzing is supposed to provide. Maintainers may assume that a project is being thoroughly fuzzed day after day, when in reality, new or changed code paths are falling out of the scope of the harness and are no longer being tested. 

In this paper, we present the first comprehensive empirical study on the degradation of fuzz harnesses in real-world projects. We analyze data from hundreds of C/\CPP projects integrated into Google's OSS-Fuzz platform and track each project's fuzzing coverage and bug finding performance from its initial OSS-Fuzz integration over several years of continuous fuzzing. In total, our study examines \data[filteredData][numProjects] projects and \data[filteredData][numHarnessChanges] different fuzz harness versions to quantify the extent of harness degradation over time. We measure how the code coverage of each harness evolves with changes to the codebase, how the rate of bugs found is affected, and conduct in-depth case studies on instances where coverage declines. Our investigation yields several notable findings. First, we find that fuzz harnesses demonstrate, on average, stable performance even as their projects evolve. The overall percentage of fuzzing coverage for most OSS-Fuzz projects tends to remain fairly stable over time, as shown in \Cref{fig:intro-plot}. This indicates remarkable longevity for harnesses as long as they continue to be successfully built. %
This is an encouraging result, as it suggests that the effort of creating a fuzz harness can pay off in the long run. We observe only a slight decline in average coverage over a period of several years, and many harnesses maintain their coverage without explicit changes.

Second, we observe high variance and individual degeneration: Although the aggregate trend is stable, we find significant variance between projects. Not all harnesses age evenly; some suffer significant losses in coverage and some surprisingly improve, depending on how the project's development interacts with the harness. In fact, the likelihood of a noticeable decline in coverage increases over time. Our analysis shows that about 5\% of fuzz targets experience a significant decline (over 5\% absolute coverage reduction) within just six months of their last update, with this percentage increasing over longer periods. We also confirm that projects with lower fuzz coverage tend to find fewer bugs in OSS-Fuzz, and that the overall level remains quite stable.

Third, we investigate the root causes of coverage drops. Using manual case studies of projects where coverage declined, we identify common causes of harness degradation. A predominant cause is project development, such as newly added features or additional input checks that are not successfully covered by the existing harness. In other cases, we found build and integration issues, e.g., changes in the build system or environment that caused certain fuzz targets to be omitted or certain features to be disabled during fuzzing, reducing coverage. We also encountered cases of irregularities in coverage measurement (e.g., coverage calculation including third-party code or incorrect mapping) that can give the appearance of a decline.
By categorizing these causes, our study sheds light on why harnesses become obsolete despite continuous fuzzing.

Finally, based on our findings, we took steps to help practitioners detect silent degeneration of the harness. More specifically, we collaborated with the OSS-Fuzz team and the open-source tool \emph{Fuzz Introspector} to implement new metrics and surface information that make potential harness issues visible to project maintainers. Such early warnings are designed to ensure that those responsible are aware of gaps in their fuzzing coverage. The goal is to avoid situations where developers believe their project is well fuzzed, when in reality fuzzing effectiveness declines without being noticed. By providing greater understanding of long-term fuzzing coverage trends and actionable insights into the reasons and semantics that cause harness degradation, we hope to make continuous fuzzing more resilient to code churn and preserve its security benefits.

In summary, we make the following key contributions:

\begin{compactitem}
\item We systematically analyze and quantify harness degradation and the effect of harness updates in OSS-Fuzz based on coverage percentage and bug-finding capability.
\item We manually analyze cases of harness coverage degradation in OSS-Fuzz and categorize them into common causes.
\item Based on this analysis, we introduce new OSS-Fuzz and Fuzz Introspector metrics and actionable insights for developers that help to proactively detect signs of harness degradation.
\item We release our complete dataset\footnote{\microtypecontext{expansion=sloppy} Dataset: \datasetDOI}, scraping code, case study notes, and analysis notebook\footnote{\microtypecontext{expansion=sloppy} Repository: \url{\repoURL}} to allow close inspection of our results and facilitate further research on this topic.
\end{compactitem}

\section{Background}

We begin by presenting background information, focusing on fuzzers and fuzzing harnesses.

Fuzzers are dynamic testing tools that execute a target program with randomized inputs. This method has established itself as an effective and practical way to detect bugs and security vulnerabilities. Modern general-purpose fuzzers generally use an evolutionary approach, where inputs are slightly modified (``mutated''). The inputs that perform well, such as reaching new parts of the code or triggering new behavior, are kept to be mutated again. To kickstart this process, it is beneficial to have a comprehensive initial set of inputs that the fuzzer can start with, which is called a \emph{seed corpus}. The set of inputs found during fuzzing is called a \emph{corpus}~\cite{li2018fuzzingSurvey, sutton2007fuzzingBruteForceBook, fuzzingbook2024}.

The term \emph{fuzzing harness} (\emph{fuzz harness} for short) is derived from software testing terminology, where a test harness describes a collection of test stubs and test drivers, which are required to execute a test suite~\cite{istqbDefinitionTestHarness}. Similarly, we use the term \emph{fuzz harness} as the collection of standardized entry points into the program under test. As test stubs are less relevant for fuzzing, we use the term \emph{fuzz target} for individual entry points~\cite{zhang2021apicraft}. 

To provide a flexible way of generating new inputs, most general-purpose fuzzers settled on a standardized function~\cite{libfuzzerDocu}. This function takes a byte vector and its length as inputs. That byte vector needs to be passed to the target program in a syntactically and semantically correct way, which is the goal of the fuzz harness. Thus, a typical fuzz harness function transforms this byte vector into data structures and calls functions of the target program. If needed, the harness includes calls to initialization and cleanup functions. So, the harness function must encode the business logic required to interact with the target application correctly. Thus, the harness needs to be updated to keep up with potential changes in the program's code base. This makes fuzz harnesses interesting potential points of failure, which we study in this paper.

OSS-Fuzz~\cite{ossFuzzRepo} is a project developed and maintained by Google that provides maintainers of open-source projects with the infrastructure to fuzz their code. To participate, open-source projects need to be integrated with the OSS-Fuzz infrastructure. This entails developing fuzz harnesses for the projects, setting up build configurations, and providing meta information~\cite{ossFuzzDocu}. To incentivize open-source projects to be integrated with OSS-Fuzz, a bounty program is offered that rewards an increase in coverage of at least 10\% or achieving a coverage rate of over 50\%~\cite{ossFuzzRewardProgram}.

OSS-Fuzz supports projects written in several programming languages and supports the x86\_64 or i386 architectures. After successful integration with OSS-Fuzz, the current version of the project is continuously built and fuzzed daily. OSS-Fuzz relies on a scalable, distributed infrastructure to use different fuzzers and sanitizers and provide automatic reporting for the project's maintainers. This includes detected bugs and detailed daily coverage results to facilitate further fuzzing~\cite{ossFuzzDocu}. To our knowledge, the OSS-Fuzz project also keeps the corpus of previous runs available for the project maintainers and uses it, or a subset, as the seed corpus for future runs. Finally, and crucially for our research, most results, such as coverage and created corpora, are publicly available data from OSS-Fuzz, which enables a long-term study such as the one conducted in this paper.

Fuzz Introspector~\cite{fuzzIntrospectorRepo} is a companion project to OSS-Fuzz that analyzes the fuzzing performance of projects in OSS-Fuzz; however, it is also possible to use it for individual projects. Fuzz Introspector provides information such as the reachable code, the reachable code complexity, and specific functions that can not be sufficiently explored by fuzzers. In addition, a web interface is provided listing all projects in OSS-Fuzz with project-specific overviews~\cite{fuzzIntrospectorProjectsOverview}.

\section{Fuzz Harness Degradation}

With the background established, we now move on to the core part of this paper, the data collection and analysis.
To evaluate the impact of fuzz harness degradation, we gather data on two central metrics of fuzzer evaluation, code coverage and bug-finding capability~\cite{klees2018evaluatingFuzzers, schloegel2024sokFuzzcrime}.
In particular, we are interested in harness performance over time, where we consider these metrics over different harness versions as a project evolves.
Furthermore, we aim to identify the reasons for harness degradation.
To this end, we manually perform case studies on code coverage drops and classify common causes.
In total, we investigate the following four research questions:

\newcommand{\rqUpdates}{RQ1: What are the effects of harness updates?}
\newcommand{\rqDegradeCov}{RQ2: Does coverage degrade if harnesses are not updated?}
\newcommand{\rqDegradeBug}{RQ3: Does the bug-finding capability of harnesses degrade if they are not updated?}%
\newcommand{\rqDropCauses}{RQ4: What are common causes for coverage degradation?}

\begin{itemize}
    \item \rqUpdates 
    \item \rqDegradeCov
    \item \rqDegradeBug 
    \item \rqDropCauses 
\end{itemize}

\subsection{Data}\label{sec:data}

To answer these research questions, we require data that covers large parts of the lifespan of fuzz harnesses, optimally containing many fuzz harnesses and versions.
Additionally, we need bug data and the availability of artifacts to study individual cases of harness degradation.
To our knowledge, OSS-Fuzz is the only project that meets these criteria.
Thus, to investigate our research questions, we collected and pre-processed several datasets related to OSS-Fuzz to create one cohesive dataset\footnote{\microtypecontext{expansion=sloppy} Dataset: \datasetDOI} as follows.
In total, we collected data from October 2016 to October 2024.

\begin{wraptable}{tr}{8.4cm}
\centering
\caption{Data points for each step of the data preparation phase.}\label{tab:datapoints}
\begin{tabular}{ l r r r }
                                 & \markerRaw                                 & \markerCleaned                             & \markerFiltered                              \\ \toprule
 \circled{1} Projects            & \data[uncleanData][numProjects]            & \data[cleanedData][numProjects]            & \data[filteredData][numProjects]        \\ \midrule
 \circled{2} Coverage            & \data[uncleanData][numCoverage]            & \data[cleanedData][numCoverage]            & \data[filteredData][numCoverage]        \\
 \circled{3} Commits             & \data[uncleanData][numCommits]             & \data[cleanedData][numCommits]             & \data[filteredData][numCommits]         \\
 \circled{4} Harness Changes     & \data[uncleanData][numHarnessChanges]      & \data[cleanedData][numHarnessChanges]      & \data[filteredData][numHarnessChanges] \\
 \circled{5} Monorail Bugs       & \data[uncleanData][numMonorailBugs]        & \data[cleanedData][numMonorailBugs]        & \data[filteredData][numMonorailBugs]   \\
\end{tabular}
\end{wraptable}

We prepared each data source in three steps, corresponding to the three columns in \Cref{tab:datapoints}. First, we collected all available data in the \markerRaw~step. Next, we removed all unusable or irrelevant data in the \markerCleaned~step. This includes data outside the period between a project's initial inclusion in OSS-Fuzz and one week after its last commit, and other cleanups, which are explained individually below. Finally, we filtered out projects with insufficient data in the \markerFiltered~step --- we require at least one coverage measurement, harness update, and bug found, as well as at least 5,000 changed lines in the project. Beyond these general steps, data source-specific actions are described individually below. The number of data points after each step can be found in Table~\ref{tab:datapoints}.

\subsubsection{Projects}\label{sec:projects} \circled{\normalfont 1} 
This shows the number of projects for each step.  Note that only C/\CPP projects are used in our dataset.

\subsubsection{Coverage}\label{sec:data-coverage} \circled{\normalfont 2} 
This contains the daily coverage reports per project and harness. \markerRaw~OSS-Fuzz publishes one coverage report per day per project and harness. We did not count a data point if no data was available on a particular day. However, we tracked the absence of data. This data was collected from the HTML coverage report\footnote{ \smaller \microtypecontext{expansion=sloppy} \nolinkurl{https://storage.googleapis.com/oss-fuzz-coverage/<project>/reports/<date>/linux/file_view_index.html}\\\hspace*{1em}e.g.:\url{https://storage.googleapis.com/oss-fuzz-coverage/sudoers/reports/20250226/linux/file_view_index.html}}, which contains line, function, and region coverage as provided by Clang's source-based code coverage feature~\cite{clangSourceCodeCoverage}. \markerCleaned~We removed around 20k additional empty coverage reports created prior to a project's initial inclusion in OSS-Fuzz.

The total number of lines for a harness is the union of all lines -- after dead‑code elimination -- that can be reached from any of the fuzz targets. This number is used in coverage measurement as the baseline for all relative coverage measures. Note that this number does not necessarily correspond with the real total number of lines in a project, as this would require all projects to turn off dead-code elimination as part of the build process, which is currently not given~\cite{ossFuzzCoverageBaselining}. This limitation in the data can cause larger jumps in coverage percentages when changing the harness alone. Thus, this causes relative coverage measures (in percentages) to be more representative of the fuzzer's performance than of the overall harnessing effort. As for our evaluation, we get similar results irrespective of whether we look at the relative coverage percentage or the total number of lines covered.

\subsubsection{Commits}\label{sec:data-commits} \circled{\normalfont 3} 
This contains the commits to the default branch. \markerRaw~We cloned every git project based on the URL provided in the OSS-Fuzz \texttt{project.yaml} under the key \texttt{main\_repo}. If the repository could not be cloned, we excluded it from our analysis, which is also the case for non-git-based repositories and projects that specify a GitHub organization instead of a repository. In total, we gathered commit data for \data[commitProjs][unclean] projects. We collected every commit to the project's default branch, including the number of added and removed lines per file, as provided by git. Some projects consist of multiple repositories that include dependencies and sometimes fuzzing-related code. Commits to these separate repositories are not included in the commit count of a project, as this would require analyzing the \texttt{Dockerfile} and differentiating between libraries and project code that happens to be in a separate repository. \markerCleaned~We removed all commits that happened before the project was added to OSS-Fuzz or after projects left OSS-Fuzz.

Note that for our evaluation, we use the number of changed lines of a project as the project's code churn or to calculate a code churn-adjusted bug finding rate. When we speak of changed lines, we count the number of all added lines as provided by git, which includes additions as well as changed lines, in files with extensions that are C/\CPP related (\texttt{.c, .h, .cpp, .hpp, .cc, .hh, .cxx, .h++, .c++, .h++, .C}).

\subsubsection{Harness Changes}\label{sec:data-harness-changes} \circled{\normalfont 4} 
These include all commits that are classified as a harness change. \markerRaw~We used heuristics based on the commits in OSS-Fuzz and the project's repository to classify the commits that introduce or update fuzzer harnesses. If the commit was part of OSS-Fuzz, we included all commits that changed a file located in the corresponding project's folder. However, we excluded changes to the \texttt{project.yaml}, \texttt{Dockerfile} (or \texttt{Jenkins} file, which was used earlier) to avoid simple changes of metadata or dependency updates to count as a harness update.

For commits to the project's repository, we include changes to files with C/\CPP file extensions, where the file path included "fuzz", however, we excluded "fuzzy", as some projects use fuzzy logic. Additionally, changes to C/\CPP files containing the fuzzing functions \texttt{LLVMFuzzerTestOneInput} and \texttt{LLVMFuzzerInititalize} were also included as harness changes. This analysis is implemented with Tree Sitter~\cite{treeSitterTool}.
\markerCleaned~As for the commit data, we removed all data points outside of the project's time in OSS-Fuzz. 
\markerFiltered~We merged harness changes that occured in a three-day span to avoid counting fixes to updated harnesses. Multiple small updates were thereby counted as one, with the intent to avoid counting a single harness update multiple times. There is also a case where the earlier-mentioned heuristics can produce false positives: If the harness is part of a larger file containing normal code, these changes are counted as harness changes. As it is infeasible to automatically differentiate non-harness-related functions from harness-related support functions, it is not possible filter them out automatically. %
During our manual analysis, we detected only one project where non-harness-related functions were located in a harness file, which we filtered out.
To avoid false negatives, we manually investigated all projects with suspicious gaps in harness updates compared to commits. We detected two projects that moved fuzzing harnesses into a separate repository, which we filtered out. Additionally, three projects only contained fuzzing examples, which we filtered out as well.

\subsubsection{Monorail Bugs}\label{sec:data-monorail-bugs} \circled{\normalfont 5} 
Monorail is an issue tracker~\cite{ossFuzzIssueTracker} containing bugs found via fuzzing or errors during the build process of the project in OSS-Fuzz. Note that Monorail has been deprecated and replaced with a new issue tracker~\cite{ossFuzzIssueTracker2}, for which we do not support data gathering; our last datapoint is on the \data[lastMonorail]. \markerRaw~Three types of bugs are reported: security bugs, general bugs, and build failures (that break the project's fuzzer build on OSS-Fuzz). We also collected meta information on bugs, such as the date of the bug, labels, and comments. %
\markerCleaned~We removed all bug reports for projects that did not have harness changes. We also made sure not to include reports from before a project was integrated into OSS-Fuzz. Unexpectedly, there were 25 such reports, which we removed from the dataset. We believe that these were created for testing purposes.

\subsection{\rqUpdates} \label{sec:rq-harness-update-effect}
We hypothesized that our selection of harness updates (see \Cref{sec:data-harness-changes}), shows positive effects in terms of coverage and bug finding rates.
To investigate our hypothesis, we first took a look at the immediate effects of harness updates and compare projects with different rates of harness updates for a broader view.

\begin{figure*}[t]
    \begin{subfigure}[h]{0.42\textwidth}
        \centering
        \includegraphics[width=\linewidth]{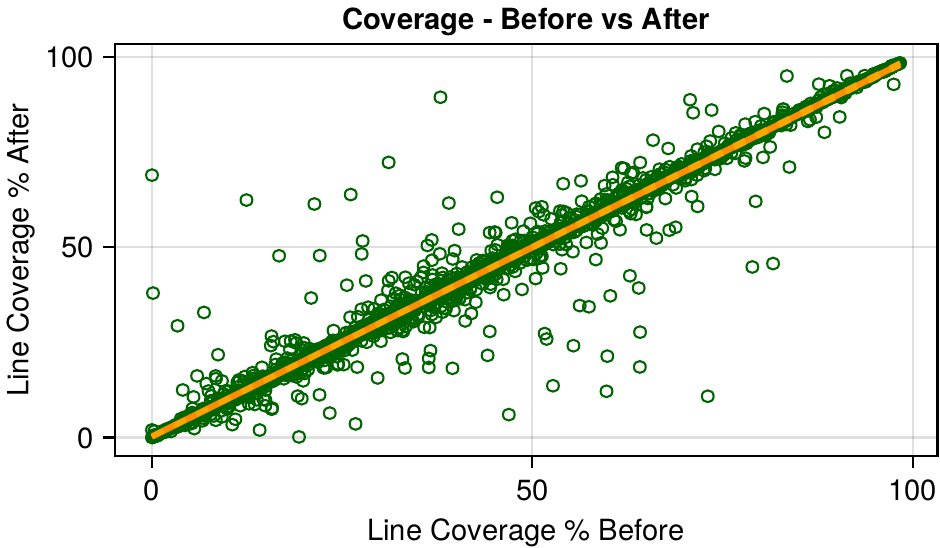}
        \caption{Coverage before and after harness updates.}
        \label{fig:harness-cov}

        \vspace{0.5em}
        
        \includegraphics[width=\linewidth]{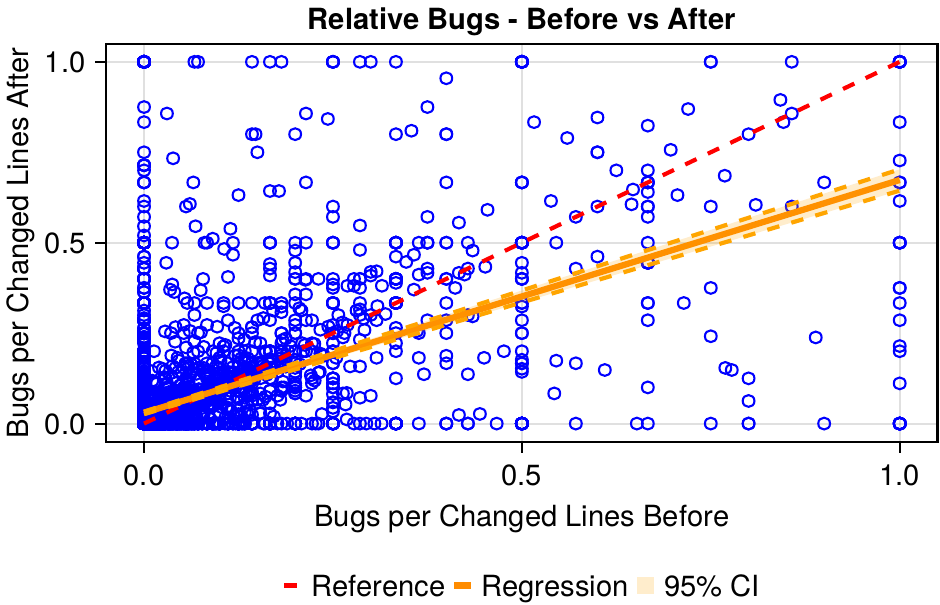}
        \caption{Bug finding before and after harness updates. Note that the vast majority of data points are located in the bottom left. We excluded spurious data above a threshold of one for legibility reasons.}
        \label{fig:harness-bug}
        
        \vspace{0.5em}
        
        \includegraphics[width=\linewidth]{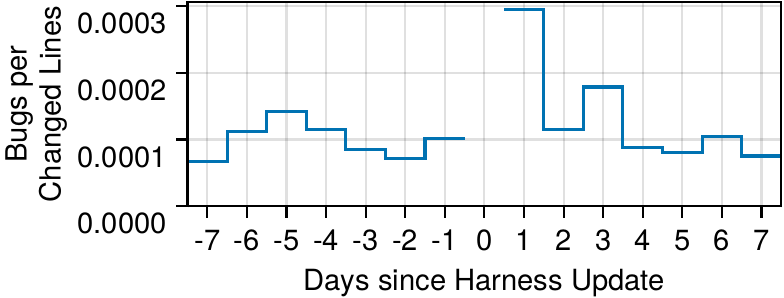}
        \caption{The immediate effect on the bug finding rate of harness updates.}
        \label{fig:harness-bug-finding}
    \end{subfigure}
    \qquad
    \begin{subfigure}[h]{0.33\textwidth}
        \centering
        \includegraphics[width=\linewidth]{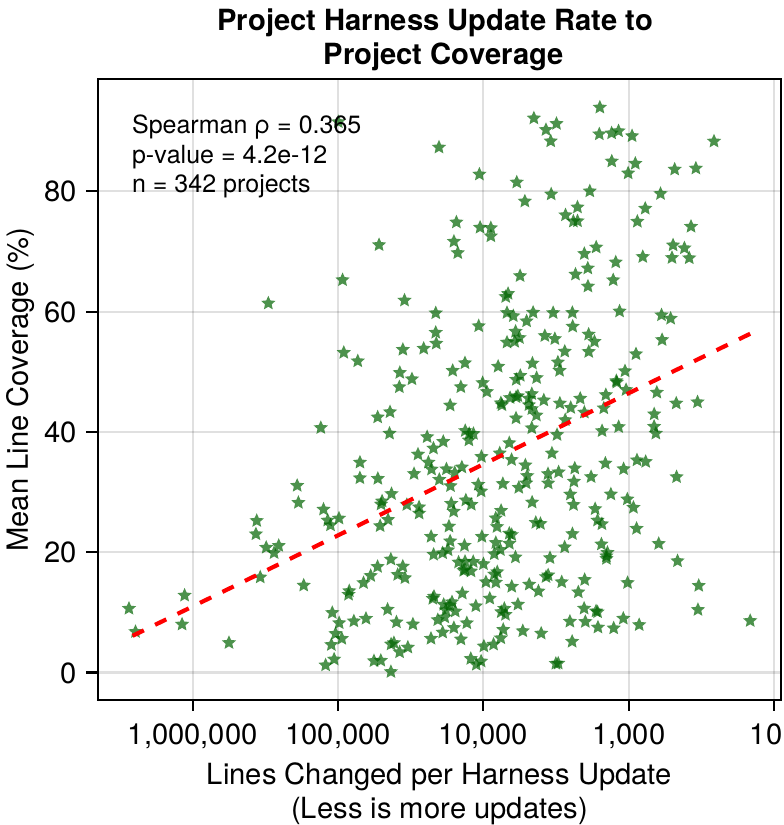}
        \caption{Correlation between harness update rate and coverage.}
        \label{fig:hur_cov_rate}
        
        \vspace{0.5em}
        
        \includegraphics[width=\linewidth]{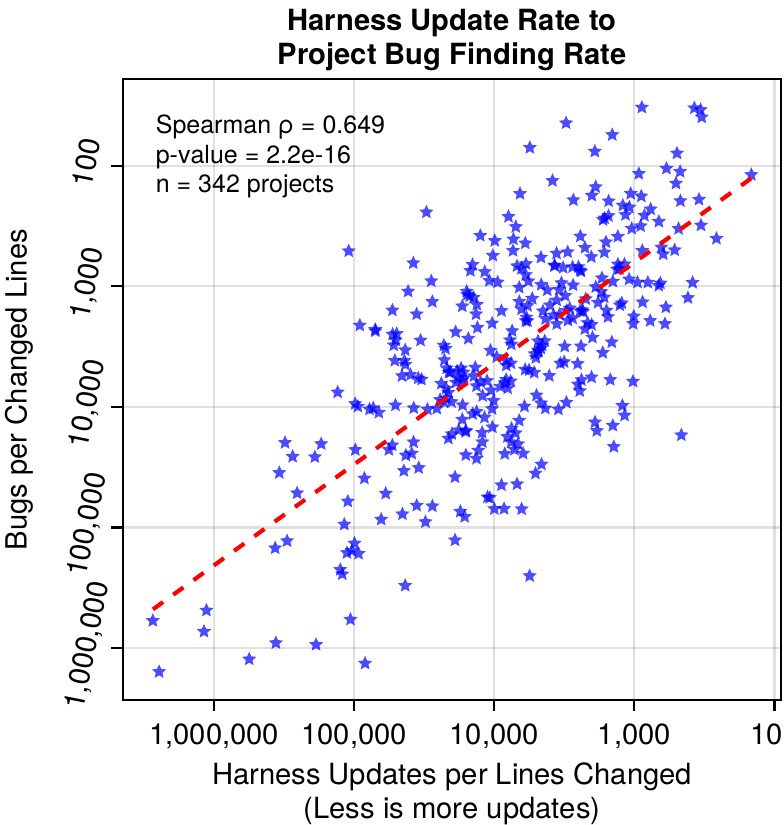}
        \caption{Correlation between harness update rate and bug finding rate. The number of lines is the cumulative sum of changed and added lines across commits.}
        \label{fig:hur_bug_rate}
    \end{subfigure}
    \caption{The effects of harness updates.}
    \label{fig:harness-effects}
\end{figure*}

Regarding the immediate effects of harness updates, we compared data on the seven days before and after a harness update. We compared the maximum line coverage in \Cref{fig:harness-cov} and the count of bugs divided by the count of changed or added lines during that time across commits (bugs per changed lines) in \Cref{fig:harness-bug}. We filtered out harness updates where another harness update occurred within the seven days and required at least one coverage measurement in the seven days before and after the harness update, which resulted in \data[rq1HarnessCount] harness versions. Additionally, we removed extreme outliers where more than one bug per changed line is found, removing \data[rq1HarnessOutliers] versions.

We performed a two-sample Wilcoxon signed-rank test for paired data on the data before and after the harness update.
For relative coverage, we got an estimated percentage change of \data[rq1Wilcoxon][perHarnessCovNoOutliers][estimate], which is statistically significant with a p-value of \data[rq1Wilcoxon][perHarnessCovNoOutliers][pValue].
This appears to be consistent with the appearance of \Cref{fig:harness-cov}.
Furthermore, to test for the sensitivity of this result on potential outliers, we performed the same analysis with data including outliers, as described in the previous paragraph.
Here, the estimated effect is \data[rq1Wilcoxon][perHarnessCov][estimate] with a p-value \data[rq1Wilcoxon][perHarnessCov][pValue], which demonstrates the robustness of our result against the treatment of outliers.
Similarly, when applying the Wilcoxon test to the harness means per project, we get an estimate of \data[rq1Wilcoxon][perProjectCovNoOutliers][estimate] with p-value \data[rq1Wilcoxon][perProjectCovNoOutliers][pValue] and, again including outliers, an estimate of \data[rq1Wilcoxon][perProjectCov][estimate] and p-value \data[rq1Wilcoxon][perProjectCov][pValue].
In summary, this demonstrates a minimal impact of harness updates on coverage.
Note that we got a similar result when we did this experiment with the total number of covered lines.

A comparison between the harness update rate (defined as the number of harness updates divided by the cumulative number of changed lines across all commits when the project was fuzzed by OSS-Fuzz) and the coverage of projects revealed a positive correlation, as shown in \Cref{fig:hur_cov_rate}.
While this is only a weak positive correlation for coverage, it is highly statistically significant,
indicating that projects that update their fuzz harnesses more often achieve higher coverage.

Beyond coverage, we further investigated whether harness updates improved the bug finding rate (the number of bugs found divided by the number of lines changed).
As before, we compared before and after using the Wilcoxon test, this time for changes in the bug finding rate.
We got an estimated improvement of \data[rq1Wilcoxon][perHarnessRelNoOutliers][estimate] with a highly statistically significant p-value of \data[rq1Wilcoxon][perHarnessRelNoOutliers][pValue].
Looking at \Cref{fig:harness-bug}, which includes a linear regression for visualization purposes, shows that most of the improvement is from improving harnesses whose bug finding rate is close to zero, where we also have the bulk of our data.
Furthermore, an analysis of sensitivity on outliers, as described above, confirms robustness of this result w.r.t. outliers with \data[rq1Wilcoxon][perHarnessRel][estimate] (p-value = \data[rq1Wilcoxon][perHarnessRel][pValue]). Analyzing the data on a per-project basis resulted in an estimate of \data[rq1Wilcoxon][perProjectRelNoOutliers][estimate] (p-value=\data[rq1Wilcoxon][perProjectRelNoOutliers][pValue]) and included outliers of \data[rq1Wilcoxon][perProjectRel][estimate] (p-value=\data[rq1Wilcoxon][perProjectRel][pValue]). In summary, this shows a significant improvement in bug finding.

Analyzing the bug finding rate per day, shown in \Cref{fig:harness-bug-finding}, we found that there is indeed a bug burst in the first days after the harness update.
This result is also confirmed when considering the correlation of each project's harness update rate to bug finding rate, as shown in \Cref{fig:hur_bug_rate}, which, other than coverage, is a strong correlation and again highly statistically significant.
We interpret these results as clear signs that harness updates have a positive effect on the bug finding rate.

\begin{boxResult}
We found strong evidence that harness updates result in immediate positive effects on bug finding, and that the frequency of harness updates strongly correlates with the bug finding rate.
Additionally, while harness updates have a neutral effect on immediate relative coverage changes, we still found a weak correlation with overall project coverage.
\end{boxResult}

\subsection{\rqDegradeCov} \label{sec:rq-degradation-code-coverage}

Now that we had studied the effects of harness updates, we investigated fuzzing performance when fuzzer harnesses were neglected. We hypothesized that if there are no more harness updates, the fuzzing coverage would degrade over time.
Note that, as we learned in \Cref{sec:rq-harness-update-effect}, harness updates had, on average, no noticeable immediate effect on relative coverage, while more harness updates correlated with a higher project coverage percentage. This indicates coverage improvements outside of harness updates.

\begin{figure*}[hbt]
    \begin{subfigure}[t]{0.49\textwidth}
        \includegraphics[width=\linewidth]{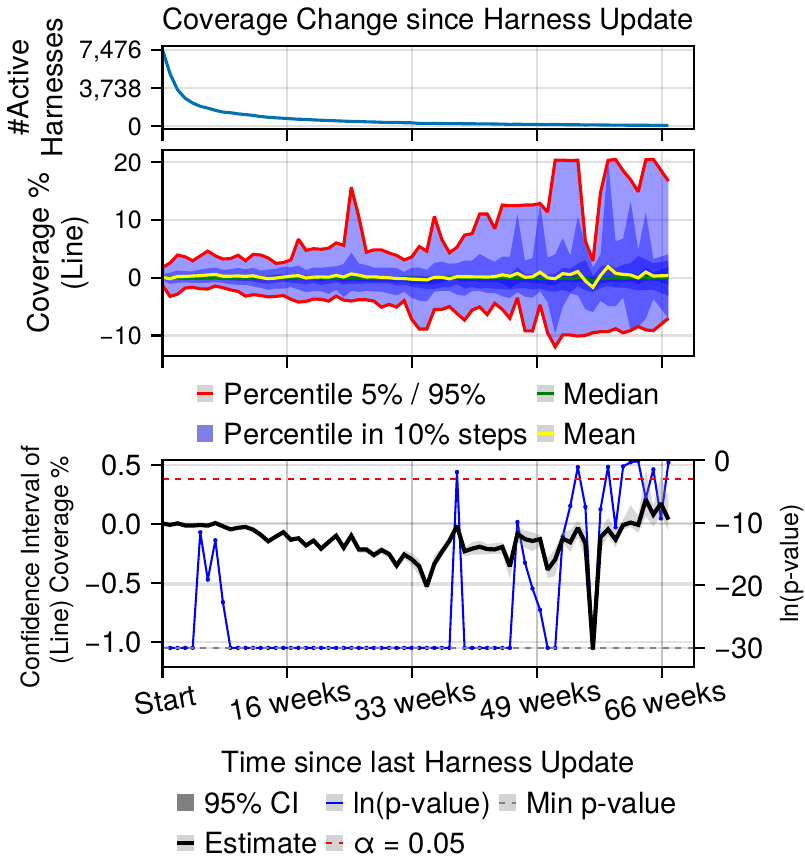}
        \caption{Harness lifetime coverage changes over time.}\label{fig:coverage-since-harness-all-time}
    \end{subfigure}
    \hfill
    \begin{subfigure}[t]{0.49\textwidth}
        \includegraphics[width=\linewidth]{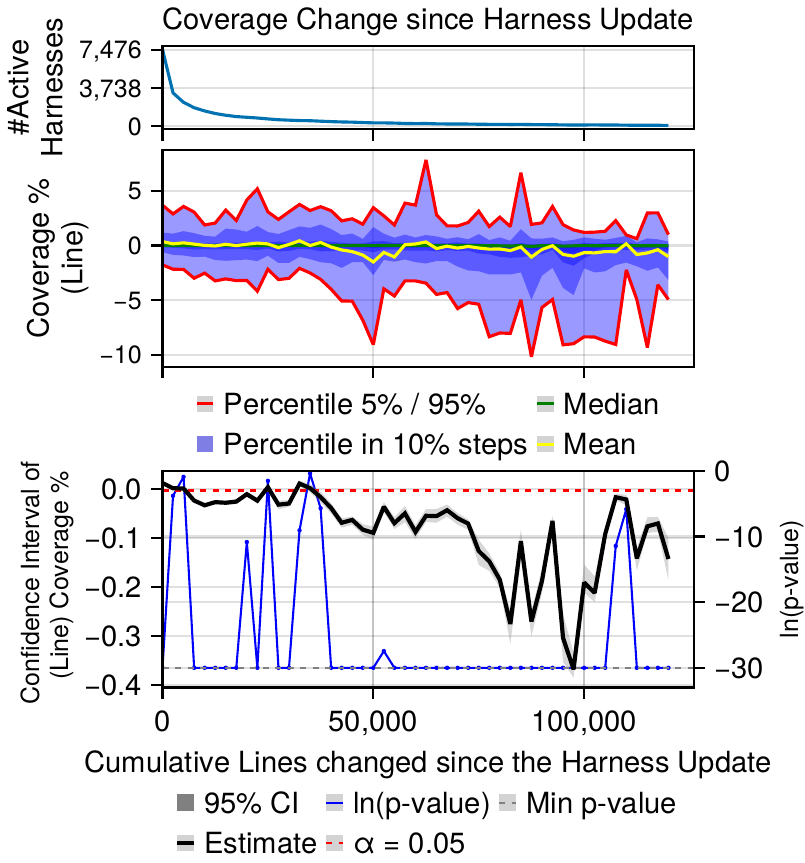}
        \caption{Harness lifetime coverage changes over churn.}\label{fig:coverage-since-harness-all-changes}
    \end{subfigure}
    \caption{Relative coverage change since a harness update across all projects. The active harnesses show the number of harnesses that have not been updated yet, representing the available data points over time.}\label{fig:coverage-since-harness-all}
\end{figure*}

To closely examine the hypothesis that coverage would degrade over time if there are no harness updates, we analyzed coverage changes during the lifetime of a harness version.
Note that we used the highest coverage found during the first three days of a harness lifetime to compensate for possibly slow coverage saturation and delay until OSS-Fuzz used the new harness version after the harness update.
We want to emphasize that we are interested in the degradation of a harness over its lifetime, not the impact of the harness update itself, which we discussed in \Cref{sec:rq-harness-update-effect}.
We exclude harnesses that completely broke and did \emph{not} count them as zero coverage, which would have imposed a substantial bias on the results.
Note that harness build failures have already been investigated by \citeauthor{nourrymyFuzzingBuildFails}, who found that around 12\% of builds are broken over time~\cite{nourrymyFuzzingBuildFails}. As a comparison, for the \data[filteredData][numProjects] projects in our final data set \data[coverageReportsMissing]\% of the daily coverage reports were not available. While this could have been caused by intermittent errors or errors in OSS-Fuzz itself, most cases were due to broken builds.

We show the overall relative coverage results over time in \Cref{fig:coverage-since-harness-all-time}. 
The top panel shows the number of harnesses; we stop showing data when dropping below \data[harnessCutoff] harnesses.
The middle panel shows the coverage change compared to the beginning of the harness update.
We see that the mean and median of the coverage over time is surprisingly stable. This means, that although harness degradation is a critical phenomenon in certain cases and should therefore be considered by developers, the negative impacts of harness degradation are on average, in OSS-Fuzz projects, shadowed by other, stronger effects. 

However, there were harness versions with higher variance of coverage decreases but also, quite surprisingly, improvements.
From our initial investigation, these improvements stemmed from bug fixes that unlock previously blocked parts of the code, and from new code that could already be covered by existing harnesses. We leave a more thorough investigation to future research, as for this work we are concerned with harness degradation and not improvements.

In addition to these summary statistics, we performed one-sample Wilcoxon tests for the null hypothesis of a zero mean of the coverage change. The last panel in \Cref{fig:coverage-since-harness-all-time} shows these p-values and associated confidence intervals for the mean coverage change.
We see a slight coverage decrease during the first 36 weeks, after which it seems that we are getting less consistent data, as we are reaching the limits of our data. Overall, there is indeed a sign of degradation, albeit quite small.

While this result paints a clear picture of small coverage degradation over time, it may not be representative, as projects do not linearly degrade over time, but rather when the project code is actually changed, which we refer to as code churn.
We investigate this alternate view in \Cref{fig:coverage-since-harness-all-changes}.
As before, we show the number of harnesses, code coverage change, and the result of the one-sample Wilcoxon tests for the null hypothesis of a zero mean coverage change.
Similarly to the time-based view, we can see a very stable mean and median, with some harness versions that exhibit higher variance.
The Wilcoxon tests also show a mostly similar result, except that we can see a small coverage decrease starting at around 75,000 changed lines.
After that, results are more sporadic but still indicate a downward trend, though we are again reaching the limits of our data.
In summary, this clearly shows that regarding harness degradation, it depends on the semantics of the actual changes of code, since not all code churn had the same impact. We further investigate this in \Cref{sec:rq-factors-harness-degradation}.

\begin{boxResult}
The coverage over time and over project churn is surprisingly stable during the lifetime of a harness version; this is mainly due to the large number of projects with stable coverage.
Overall, we find a statistically significant but small reduction in coverage if harnesses are not updated.
However, individual projects can still have large decreases or increases in coverage. This indicates that not time or code churn itself is the primary driver of harness degradation, but the semantics of the code changes.
\end{boxResult}

\begin{figure*}[t]
    \begin{subfigure}[t]{0.49\textwidth}
        \centering
        \includegraphics[width=\linewidth]{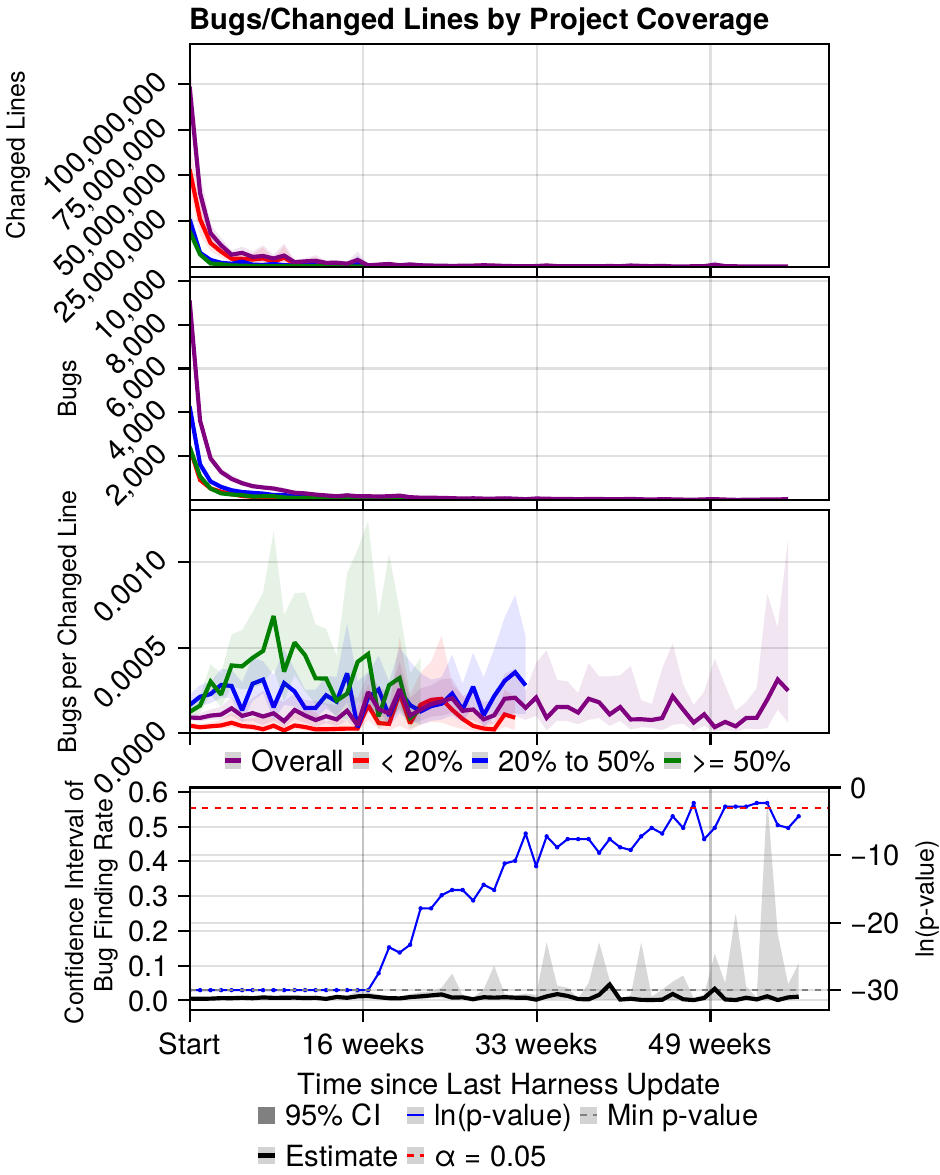}
        \caption{Over time.}\label{fig:bugs_per_lines_coverage_time}
    \end{subfigure}
    \hfill
    \begin{subfigure}[t]{0.49\textwidth}
        \centering
        \includegraphics[width=\linewidth]{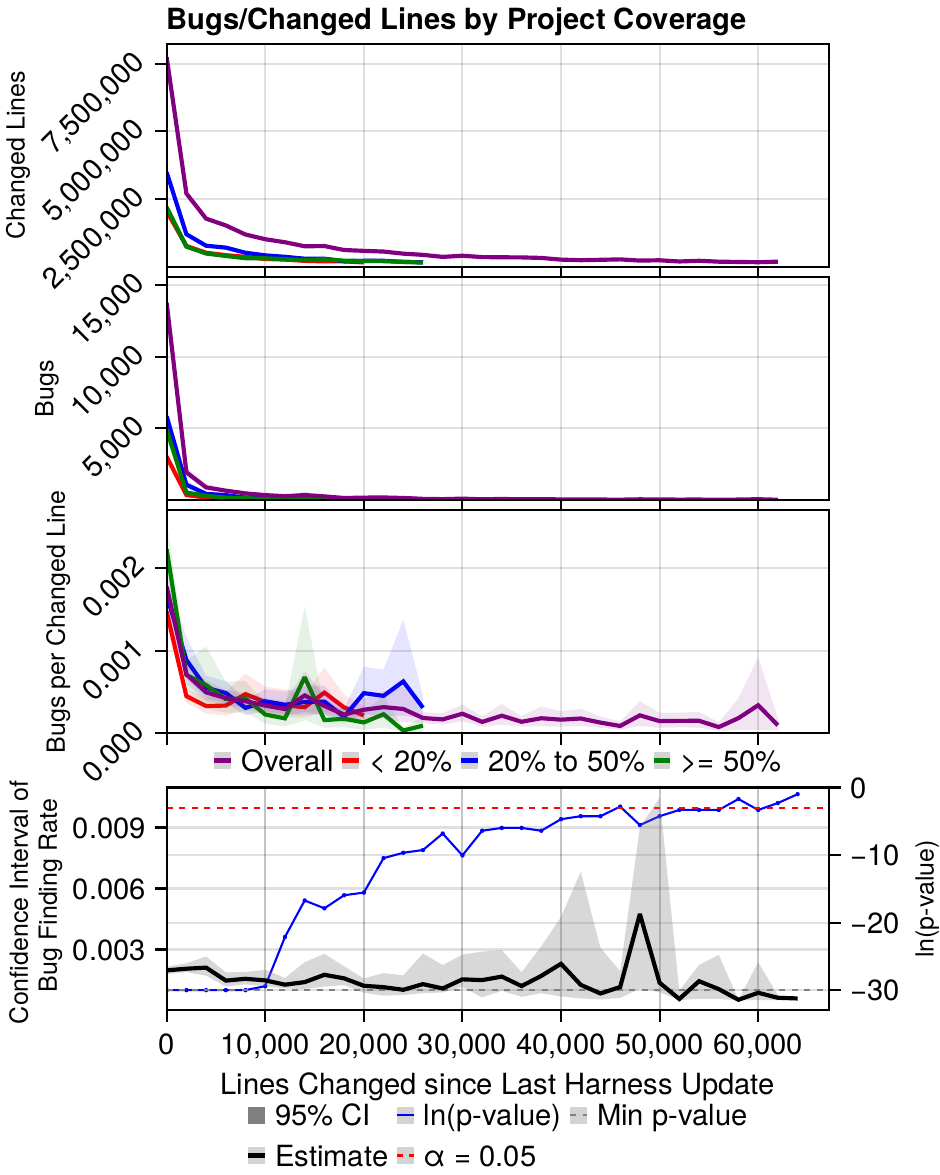}
        \caption{Over project code churn.}\label{fig:bugs_per_lines_coverage_lines}
    \end{subfigure}
    \caption{Bug finding rate for projects with under 20\%, between 20\% and 50\%, and over 50\% coverage.}\label{fig:bugs_per_changed_loc}
\end{figure*}

\subsection{\rqDegradeBug} \label{sec:rq-degradation-bug-finding}

\begin{wrapfigure}{r}{0.35\textwidth}
        \centering
        \includegraphics[width=\linewidth]{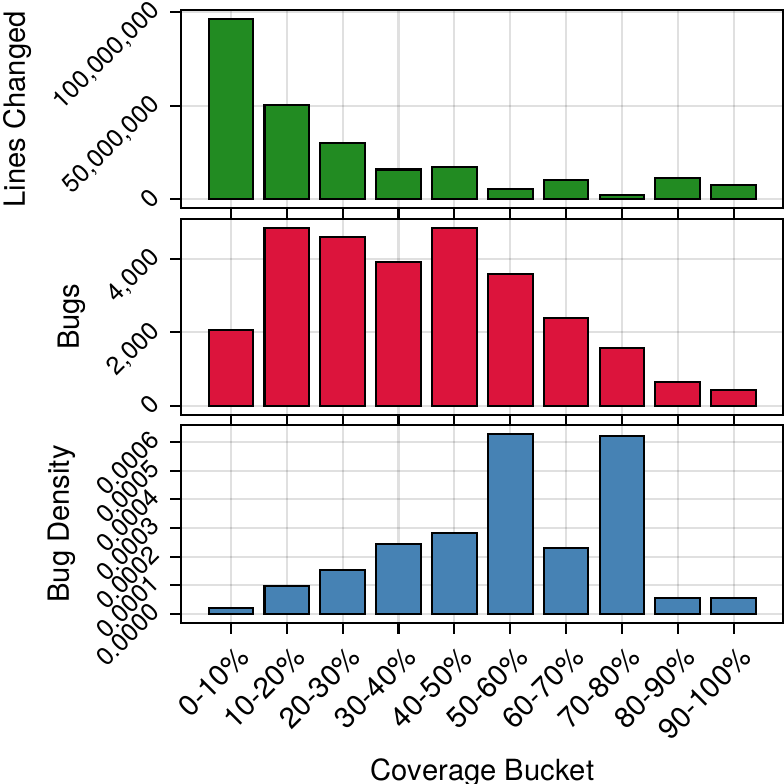}
        \caption{Overall harness data grouped by 10\% coverage buckets.}\label{fig:cov_buckets}
\end{wrapfigure}

Similar to coverage, we hypothesized that fuzzers lose their capability to find bugs if harnesses were not updated.
To investigate this hypothesis, we analyzed the bug finding rate over the lifetime of a harness version.
We visualize this data in \Cref{fig:bugs_per_lines_coverage_time}, with the first three panels showing the number of changed lines, found bugs, and the ratio of bugs found per changed lines (bug finding rate).
Note that we used a bootstrapping approach, randomly sampling results for harnesses repeated \data[bootstrapReps] times, to provide 95\% confidence intervals for these three panels.
We also cut off data once we reached less than \data[harnessCutoff] available harnesses per group.
As previously mentioned, these plots are relative to the time of the last harness update.
Additionally, we again removed outliers where more than one bug is found per changed line.
Looking at the "Overall" result, we can see that there is an initial bug burst after harnesses are updated, at least in terms of the absolute number of bugs. However, this initial bug burst also correlates with a code churn burst, clearly seen in the third panel. When calculating the ratio of bugs per changed lines, we see a surprisingly stable rate of bug finding.
Note that the bug burst we observed in \Cref{sec:rq-harness-update-effect} would also be visible in this figure; however, these results are too zoomed out, and the initial bug burst is still below some of the later peaks of the overall result.

We performed a two-sample Wilcoxon tests on the code churn adjusted bug finding rate, relative to the bug finding rate achieved in the first week, which confirm this result.
Here we take each week per harness separately and calculate the bug finding rate.
The result can be seen in the fourth panel of \Cref{fig:bugs_per_lines_coverage_time}. It shows that the overall bug finding rate does not deviate strongly from the initial result.

As established in previous research~\cite{bohme2022reliability}, coverage has a strong correlation with the ratio of bugs found, which means that projects with good coverage are able to uncover more bugs via fuzzing.
Splitting up our results into three coverage groups, of up to 20\%, 20-50\%, and above 50\%, we aim to investigate the effect that coverage has on the bug finding rate if the harnesses are not updated. Indeed, we can confirm the previous result with our data. And more relevant to our research question, this correlation also holds true concerning the longevity of harnesses. That is, harnesses seem to mostly keep their relative bug finding capability even if they are not updated.

Perhaps surprisingly, considering that OSS-Fuzz provides a monetary reward, the number of projects with coverage over 50\% is not well represented, at least in terms of changed lines, as shown by the first panel in \Cref{fig:cov_buckets}.
Further, as can be seen in the third panel, coverage seems to correlate with bug finding rate up to 50\%, after which we get more varied results.
The cause, as far as we can tell, is simply unrepresentative data for projects with high coverage percentages.

Going beyond the bug rate change over time, we also investigated the change over project churn, that is, the number of changed lines in a project since the start of the harness.
We provide these results in \Cref{fig:bugs_per_lines_coverage_lines}.
Note that now we \emph{do} see the initial bug burst, which is followed by a quite stable bug finding rate.
For this reason,  we used the second group of results --- where each group contains \data[cumsumBuckets] lines of changed code --- for the two-sample Wilcoxon tests to avoid relating the tail of the bug finding rate against the first peak.

When comparing these results against our results for the bug finding rate over time, we found similar results, except for the initial bug burst, and we again confirmed a very stable bug finding rate even if no harness updates are performed.
However, for the different coverage buckets, we got some differing results. While we got the expected result in the initial bug burst, that is, higher coverage has a higher bug finding rate, we got more mixed results later on, which we attribute to the limited amount of data left at this point.

\begin{boxResult}
We find that after an initial bug burst, the bug finding rate is surprisingly stable, even if harnesses are not updated. This holds over time as well as over project code churn.
We can also confirm that after this initial burst, the bug finding rate remains relatively stable over time for different coverage percentage buckets.
\end{boxResult}

\subsection{\rqDropCauses} \label{sec:rq-factors-harness-degradation}

Even though our data shows that overall C/\CPP projects in OSS-Fuzz maintain their coverage quite well, there is still great variance, meaning that a significant portion of projects show signs of degradation. Furthermore, coverage is important for a high bug detection rate.
It is therefore important to identify cases where coverage degradation happens, especially when coverage decreases without maintainers noticing.

To evaluate causes for coverage drops, we first identified relevant instances of coverage degradation. For this purpose, we filtered for coverage drops that reduced coverage by 5\% or more, compared to the monthly average before and after the drop. We chose 5\% to get a clear signal that a degradation occurred and to keep the number of cases to study manageable. 
This filter also helps us focus on cases that resulted in long-term harness degradation.
Additionally, we filtered for days with a 5\% coverage drop, which allowed us to focus the case studies on the days when the drop actually happened.
However, several projects with unstable coverage exist, where the previous two filters still retained many uninteresting cases, due to repeated fluctuations.
We removed these cases by additionally filtering for cases where the \emph{maximum} coverage of the previous month dropped by 5\% over the following month.
Note that, again, we excluded harnesses that no longer built at all.
While this can be seen as an extreme form of harness degradation --- which results in zero coverage --- it is also easily detected and thus of no further interest in this context.
This resulted in \data[numDegradationStudies] interesting instances of harness degradation, which we all studied.

\begin{figure*}[hbt]
    \centering
    \begin{subfigure}[c]{0.28\textwidth}
        \centering
        \includegraphics[width=\linewidth]{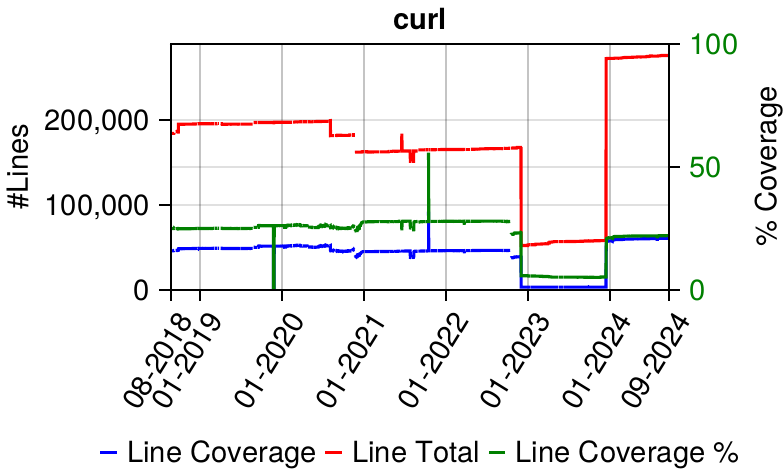}
        \caption{Curl's coverage history.}
    \end{subfigure}
    \qquad
    \begin{subfigure}[c]{0.22\textwidth}
        \centering
        \includegraphics[width=\linewidth,trim=0 0 2pt 0,clip]{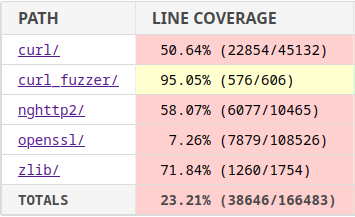}
        \caption{Before breakage (end 2022).}
    \end{subfigure}
    \qquad
    \begin{subfigure}[c]{0.22\textwidth}
        \centering
        \includegraphics[width=\linewidth,trim=0 0 16pt 0,clip]{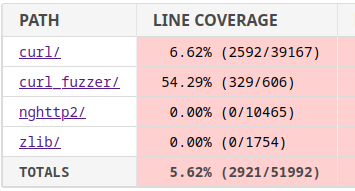}
        \caption{After breakage (end 2022).}
    \end{subfigure}
    \caption{Reports and coverage for Curl.}\label{fig:curl-report}
\end{figure*}

We categorized all of these coverage drops following the idea of grounded theory~\cite{glaser2017discovery}.
Two judges reviewed each coverage drop, based on multiple data sources, such as comparing coverage reports on the day before and on the coverage drop. If available, this also included the Fuzz Introspector reports for these days.
Furthermore, judges are allowed to investigate additional information, such as commits, and build logs.
The judges first individually categorized each case, followed by a discussion; if needed, this process is repeated to finally settle on a common categorization for each instance.
Additionally, during our initial explorative investigation, we documented other related cases, reaching a total of \data[classificationCounts][total] case studies.
The judges decided on the following categories and results related to harness degradation.

\subsubsection{Harness Build Failure} (Cases: \data[classificationCounts][buildFailure]).
We identified multiple instances where individual fuzz targets no longer build correctly.
This can be caused by a dependency no longer being available, a misconfiguration of the compilation environment (such as compiler flags), or a variety of similar reasons that have been studied in more depth before~\cite{nourry2023human, nourrymyFuzzingBuildFails}.
Note that in contrast to previous work, the cases we investigated do not lead to a full but only a partial build failure, as only some fuzz targets no longer build, which still severely impacts fuzzer performance.
These kinds of build failures usually surface as a small reduction in the total number of covered lines but a large drop in the coverage percentage. Note that these cases are sometimes not reported to Monorail, as they are typically configuration mistakes or errors handled by the build script, so no error is surfaced; thus, a build failure cannot be reported.

For example, Curl~\cite{curl_github} experienced around one year of ineffective fuzzing (see \Cref{fig:curl-report}) which started at the end of 2022~\cite{ossFuzzCurlCovBefore} and lasted until the issue was resolved in early 2024~\cite{ossFuzzCurlCovAfter}. The reason was that OpenSSL~\cite{openssl_github} was no longer building~\cite{ossFuzzCurlCovBug}. However, that did not completely stop the build process, as some fuzz targets were still working.

\subsubsection{Project Code Added and Code Churn} (Code added: \data[classificationCounts][projectCode], code churn: \data[classificationCounts][codeChurn]).
In these cases, the project is developed, the code size is increased (code added) or semantics are changed (code churn).
However, the harnesses are not updated to reflect these changes appropriately.
This is what we initially imagined as the typical reason for harness degradation.
However, many projects in OSS-Fuzz are quite mature, and this effect can mainly be observed in projects with active development, which leads to a relatively low number of observed cases.
This degradation usually surfaces either as an increase in the total number of lines of covered code accompanied by a drop in the percentage of code coverage or as a coverage drop only.

As this category is at the core of harness degradation as originally aimed for by our study, we further explore the causes that lead to this kind of harness degradation. We identified the following types of root causes that can lead to harness degradation.

\paragraph{Features Added} (Cases: \data[churnCauseCounts][features]).
An expected reason for harness degradation due to code churn is the addition of new features. We encountered multiple cases where significant new features were added to the code base. If fuzzers fail to find inputs to cover this new code or the code is not reachable by existing harnesses, then fuzzing performance degrades.
Typical for this group are many smaller additions of code, which we saw multiple times in Solidity~\cite{ossFuzzCovReportsSolidity}. Alternatively, one large addition is made, as seen, for example, in Wuffs~\cite{ossFuzzCovReportsWuffs}.
Additionally, even performance-only improvements can have a negative impact on coverage, as we observed, multiple cases, such as in Apache Arrow~\cite{ossFuzzCovReportsArrow}, where a larger number of lines implement CPU-specific code variants, for example, alternative implementations using SIMD instructions. However, these alternative code paths are not fuzzed, as it is common to use automatic variant selection based on available CPU features.

\paragraph{Stricter Checks} (Cases: \data[churnCauseCounts][fuzzerRegression]).
In these cases, existing checks are made stricter or additional checks are added, which makes it more difficult for fuzzers to cover the code.
For example, stricter input parsing or, in other words, restricting the allowed input space, makes it harder for fuzzers to cover code and often requires rebuilding the input corpus as the expected format changes.
This is also commonly caused by changes to the program logic, especially if the required conditions to execute functionality are restricted.
These changes only have a large impact on coverage if they affect a central or early part of the code. As this is usually difficult for projects to change, the number of cases in our case studies is rather small. However, keep in mind that during our case studies we only investigated large coverage jumps, and we expect this type of degradation to be more common, albeit with a smaller impact. A notable example happened in open62541. Here, a change to the message processing loop led to early stoppage when a specific message type is received. However, the fuzzer corpus had many inputs where interesting functionality only happened after one of these messages was received, thus causing the existing corpus to become substantially less effective and making it more difficult for the fuzzer to create viable inputs~\cite{ossFuzzCovReportsOpen62541}. 

\begin{wrapfigure}{r}{0.4\textwidth}
    \includegraphics[width=\linewidth]{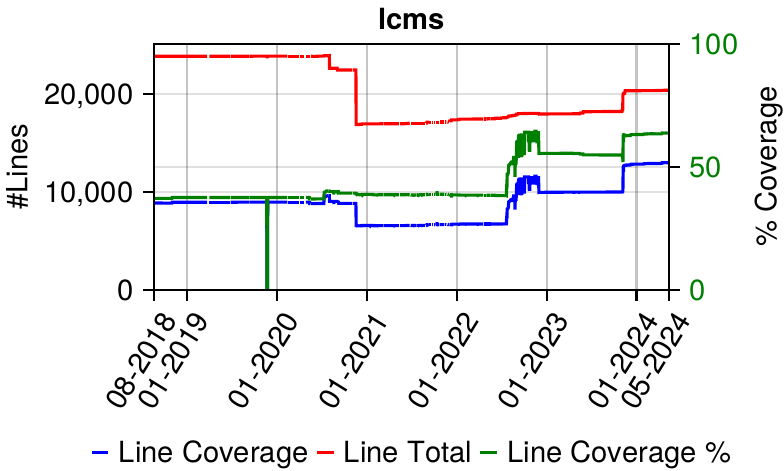}
    \caption{Little-CMS's coverage history.}
    \label{fig:lcms-cov}
\end{wrapfigure}

\paragraph{Mistakes During Harness Updates} (Cases: \data[churnCauseCounts][harnessFail]).
While also not exactly common in our data set, another reason for coverage degradation due to code churn is mistakes when updating harnesses.
This encompasses all cases where a change of one or several harnesses resulted in an immediate and permanent decrease in the absolute number of covered lines.
An example of this is a change to the Little-CMS project, a small-footprint color management engine, at the end of 2022, where the developers, according to commit messages, aimed for a simplification of the harness, which led to decreased coverage, as can be seen in \Cref{fig:lcms-cov}.

\paragraph{Other}  (Cases: \data[churnCauseCounts][other]).
Finally, we encountered one case where non-determinism in a reworked fuzzer harness actually caused severe degradation, as many inputs were wrongly reported as size zero, causing them to be removed from the seed corpus~\cite{ossFuzzCovReportsZydis}.

Additionally, there were two cases where the possible causes were too complicated for us to judge, as they encompassed numerous commits. They likely fall into the case of added features, but had very little added code or stricter checks, and we could not verify this exact cause.
\\

\noindent The aforementioned events are, according to our data set, typical of harness degradation due to code churn and can therefore also be interpreted as signals for developers that their harness might need an update. We therefore recommend that developers either get alerts on harness degradation or review the harness performance when a larger number of new features or big features are added (e.g., at version jumps). Other events that likely impact coverage include changes of parsing or validation of inputs, or changes to processing logic. Furthermore, we encourage developers to verify that any changes to the harness do not impede the effectiveness of the fuzzer. We expand on these recommendations in more detail in ~\Cref{sec:recommendations-oss-fuzz}.

\paragraph{Reactivation of Previously Eliminated Dead Code} (Cases: \data[churnCauseCounts][revival]).
While the aforementioned causes for code churn-based harness degradation can be related to specific events that negatively affected code coverage, we also encountered another type of phenomenon. We identified several cases where the relative coverage of the project decreased while code was changed or added. However, in these cases, the code churn itself was not driving the relative coverage degradation. Instead, the changes caused the compiler to no longer eliminate previously eliminated dead code. If the absolute number of potentially reachable code lines rises, and if covered lines do not also rise proportionally, then the coverage percentage decreases. Note that this is not a negative behavior per se. Thus, we do not consider this an actionable event for developers.

The exact reasons for the compiler to no longer eliminate code are varied, though this is generally caused by changes to the harness setup, adding functionality, or restructuring code. %
This also explains how some harness updates can actually decrease the relative coverage, even though they benefit the actual fuzzing process in practice.

\begin{figure*}[th]
    \centering
    \begin{subfigure}[c]{0.30\textwidth}
        \includegraphics[width=\linewidth]{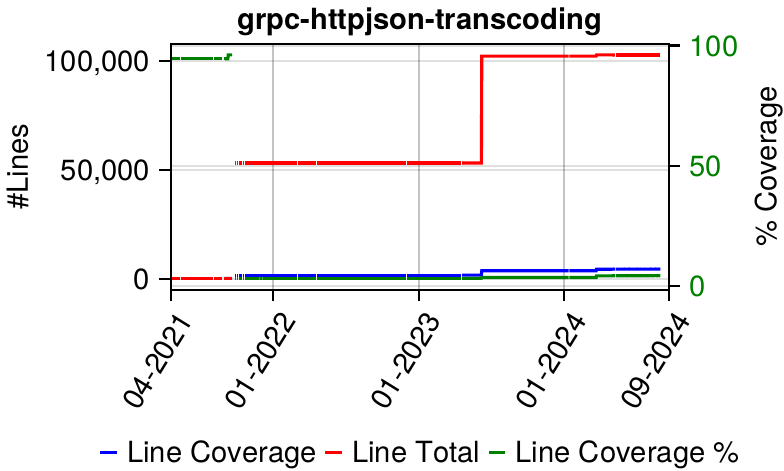}
        \caption{Coverage over time.}
        \label{fig:grpc-cov}
    \end{subfigure}
    \hfill
    \begin{subfigure}[c]{0.22\textwidth}
        \centering
        \includegraphics[width=\linewidth,trim=0 0 50 0,clip]{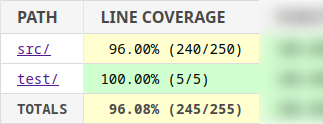}
        \caption{Initial report.}
        \label{fig:grpc-initial}
    \end{subfigure}
    \hfill
    \begin{subfigure}[c]{0.22\textwidth}
        \includegraphics[width=\linewidth,trim=0 0 5pt 0,clip]{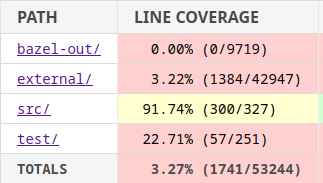}
        \caption{Adding external code.}
        \label{fig:grpc-add}
    \end{subfigure}
    \hfill
    \begin{subfigure}[c]{0.22\textwidth}
        \includegraphics[width=\linewidth]{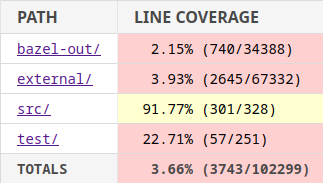}
        \caption{Adding more external code.}
        \label{fig:grpc-add-again}
    \end{subfigure}%
    \caption{The coverage performance and report of grpc-httpjson-transcoding.}\label{fig:grpc}
\end{figure*}

\subsubsection{External Code in Coverage Report} (Cases: \data[classificationCounts][libraryCode]).
Within this category, we classify projects that include code that is not part of the core project in the coverage report, such as external libraries or third-party code.
This is important for coverage measurement, which is a potentially error-prone process~\cite{schloegel2024sokFuzzcrime}, as this additional code can distort the coverage measurement of the actual program under test.
However, it is not quite clear if external code should be included in the coverage report or not.
A reason for not including external code is simply to focus on the achieved coverage of the target subject, which is, after all, what the maintainers of the project have the most influence over.
Another argument can be made that if the external code is already fuzzed separately, it might be less interesting to include this code in the coverage report.
However, some projects \emph{depend} on external code, without which they could not function, and thus, this code can be more interesting to include in the coverage report.
Moreover, there is no guarantee that this external code is fuzzed or fuzzed in a way that is representative of the usage from this specific project.
To keep things short, there is a case to be made for viewing coverage under different lenses.
However, for this study, external code should not be included in the coverage report to get a clear picture of the harnessing and fuzzing efforts of the core project itself. Especially changing the amount of external code fuzzed can distort coverage over time measures.

An extreme example is a 255 line project~\cite{ossFuzzCovReportsGrpcHttpjsonTranscoding} (grpc-httpjson-transcoding~\cite{grpc_httpjson_github}), shown in \Cref{fig:grpc}. Towards the end of 2021, the Protobuf~\cite{protobuf_github} and googletest\cite{googletest_github} libraries with 42k lines of code~\cite{ossFuzzCovReportsGrpcHttpjsonTranscoding2} were included in the coverage report of this project, as can be seen in \Cref{fig:grpc-add}. The resulting drop reduced coverage by 93\%. This happened again in mid-2023~\cite{ossFuzzCovReportsGrpcHttpjsonTranscoding3}, adding another \textasciitilde 50k lines of code, as shown in \Cref{fig:grpc-add-again}. Overall, any nuance of the project's core coverage was lost due to including these external libraries, as shown in \Cref{fig:grpc-cov}.

These cases can be observed by a large relative change in total lines covered and a drop in coverage percentage. While the core project code can remain well-fuzzed, it will be difficult to detect harness degradation of the core project, as the coverage report is (predominantly) influenced and distorted by the external code.

\subsubsection{Coverage Errors: Corpus Size Decrease and Low Corpus Size} \label{sec:low-corpus-factor} 
\begin{wrapfigure}{r}{0.4\textwidth}
    \includegraphics[width=\linewidth]{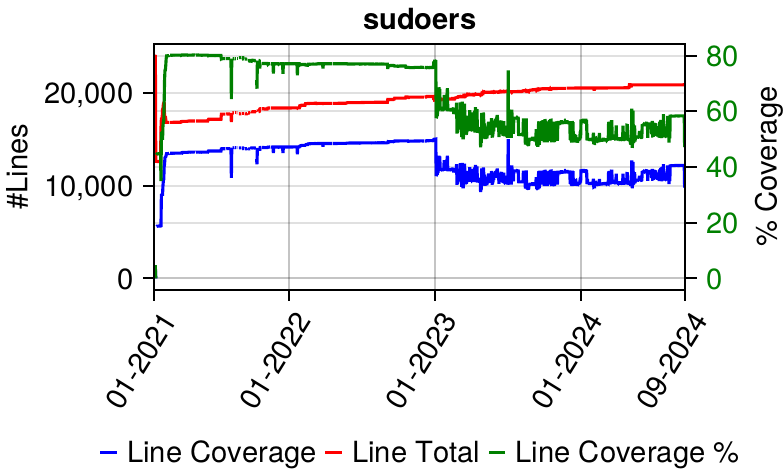}
    \caption{Sudoers coverage history.}
    \label{fig:sudoers-cov}
\end{wrapfigure}

(Size decreases: \data[classificationCounts][entriesDecrease], low size: \data[classificationCounts][entriesLow]).
We noticed several cases where the reported corpus size created by the fuzzer dropped by several magnitudes.
For example, one harness for Sudo~\cite{sudo_github} decreased from \textasciitilde 12k corpus entries~\cite{ossFuzzCovReportsSudoers} to 21~\cite{ossFuzzCovReportsSudoers2} at the start of 2023, as shown in \Cref{fig:sudoers-cov}.
In general, once this drop happens, a small corpus size and unstable coverage percentage will remain, see for example, the Sudoers project.

We investigated the code related to the coverage measurement and identified that, for projects with this phenomenon, libFuzzer either detects an OOM (2GB worker memory), or a coverage measurement times out.
However, OSS-Fuzz simply ignored these incidents, causing the coverage measurement to be reported as successful while only partially completing the coverage run.
We have since worked with the OSS-Fuzz team to fix this behavior.
Note that in cases where the fuzzer only produces a tiny corpus, it is likely that fuzzing is ineffective.

For example, consider a project with a seed corpus that covers 80\% of the project.
However, the project uses a data integrity check, stopping the fuzzer from creating new accepted mutated inputs.
This project will pass the required coverage measurements based on the seed corpus alone, but the fuzzing process is ineffective, and no interesting inputs beyond the seed corpus will be found.
As such, the reported corpus size of fuzzers is an interesting metric.

\subsubsection{Others} \label{sec:others}

We encountered intermittent coverage~\cite{clangSourceCodeCoverage} errors (\data[classificationCounts][gcovError]), causing nonsensical coverage reports, explaining some single-day variations in coverage. Additionally, some maintainers intentionally removed fuzz targets (\data[classificationCounts][intededHarnessRemoval]), causing a reduction in code coverage. Finally, in some cases (\data[classificationCounts][unknown]), we lack the required familiarity with a project to make a confident classification of the degradation.

\begin{boxResult}
We identify four common causes for coverage drops. Harness build failures, code churn, added external code, and errors during coverage measurement. Regarding code churn, we found several specific events that caused the coverage degradation based on our case studies. This includes the addition of new features by either one large or multiple small updates, changes to the program logic, and checks, especially when impacting input parsing. All of these events should therefore be clear signals for developers to consider a harness update. Furthermore, even though rare, we encountered few cases were mistakes during harness updates actually decreased fuzzing efficiency. Thus, when updating their harnesses, developers should make sure the changes do not decrease the absolute coverage.

\end{boxResult}

\subsection{Threats to Validity}

Finally, we reflect on possible issues and mitigations regarding generalizability, methodological mistakes, and the selection of experiments from which we draw our conclusions.

\paragraph{External validity} Refers to the degree to which our results can be generalized to fuzzing harnesses outside of our analysis. We analyzed all OSS-Fuzz projects (i) that are written in C/\CPP (ii) for which we could access their git repository (iii) and where we could detect harnesses (iv). See \Cref{sec:data} for more details. In the following, we discuss each point and how it influences generalizability.

\begin{inparaenum}[(i)]
    \item Projects accepted in OSS-Fuzz \enquote{must have a significant user base and/or be critical to the global IT infrastructure}~\cite{ossFuzzNewProjects}.
    Thus, we expect projects included in OSS-Fuzz to be open source, have an incentive for securing the project, and tend to be quite mature.
    Indeed, while not all harnesses are optimal, many project maintainers spend significant effort on integrating fuzzing.
    Many projects in OSS-Fuzz are mature and have relatively little new functionality added, as we also observed in \Cref{sec:rq-degradation-code-coverage}.
    Additionally, OSS-Fuzz offers a bounty program to incentivize integration and well-maintained harnesses~\cite{ossFuzzRewardProgram}.
    Regarding the effects on generalizability, we expect these selection criteria to favor mature projects with an incentive to create effective fuzzer harnesses.
    However, we expect that time spent on security and fuzzing by open-source maintainers depends on their priorities~\cite{wen2019empirical}.
    In theory, the project's maintainers should be motivated by the OSS-Fuzz reward program~\cite{ossFuzzRewardProgram}, but we only observed around a third of projects to actually surpass the 50\% coverage mark.
    \item While the choice of C/\CPP projects is to focus our research, we expect that generalizing to memory-safe languages will be difficult, especially regarding bug-finding capability, as this depends on available oracles, such as AddressSanitizer~\cite{serebryany2012addresssanitizer}, which are not available for all bug types.
    Also, we identified dependency management of C/\CPP as an issue that negatively influences harness maintenance, which might differ from other languages.
    \item The choice to focus on Git repositories is reasonable, as this still allowed us to get data for most projects in OSS-Fuzz (\data[numProjectsWithCommits] / \data[numProjectsWithCoverageData]).
    However, it is plausible that projects not using Git might have a different engineering culture, which could influence generalizability.
    \item Our method of detecting harness changes (see ~\Cref{sec:data-harness-changes}) likely introduces some bias.
    This is because projects that separate fuzzer harnesses from the main project presumably see the development of the fuzzing components as independent of the project’s code.
    This is an error-prone approach, as it can lead to code version mismatches and accelerate harness degradation.
    However, during a manual investigation, we only found two projects that use separate repositories, which we excluded from our data as described in \Cref{sec:data-harness-changes}.
\end{inparaenum}

\paragraph{Internal validity} Refers to the degree to which our study minimizes potential methodological mistakes.
While we cannot guarantee the absence of errors in our experiments, we follow a twofold approach to mitigate this issue.
First, we combine our quantitative analysis with manual analyses.
Second, we open-source all code for the experiments, scripts, and case study notes.\footnote{\microtypecontext{expansion=sloppy} Repository: \url{\repoURL}}

\paragraph{Construct validity} Refers to the degree to which our study measures what we intend to measure.
We address this by using multiple metrics to measure and assess harness degradation and the associated effects.
Note that we observed a systematic coverage measurement error in the OSS-Fuzz infrastructure, as described in \Cref{sec:low-corpus-factor}, as well as a few sporadic coverage measurement errors during our evaluation of the case studies, as described in \Cref{sec:others}. Fuzzing harnesses can also suffer from non-determinism, which we only sporadically observed in our case studies, but as there is currently no mechanism to detect this in OSS-Fuzz, we cannot rule out that it is a factor.

\subsection{On the Longevity of Fuzz Harnesses}
With the evaluation and case studies done, our data shows that fuzz harnesses in OSS-Fuzz on average show a remarkable longevity. The coverage over time and over project churn is surprisingly stable during the lifetime of a harness version, and the coverage reduction is statistically significant but minor. Similarly, after an initial burst, the bug-finding rate is also stable.
While we consider this a surprising and positive finding for the practice of fuzzing, it does not mean that updating harnesses is superfluous. Our data shows high variance and many examples of projects suffering from harness degradation even after a long stable phase, as indicated in~\Cref{fig:intro-plot}.
Still, the following question remains: Why do some projects remain stable for years while others experience abrupt failures? The simple answer is: Projects remain stable when there are no changes that adversely affect fuzzing performance. %
It is difficult to predict fuzzer performance on an abstract level, as performance depends on the project, the harnesses, the used fuzzer, and especially the semantics of the changes made. Based on our case studies, we identified common causes for coverage drops, as described in \Cref{sec:rq-factors-harness-degradation}, of which developers should be especially mindful. However, we advise developers to set up automated monitoring of the performance of the fuzzing setup, as degradation can be sporadic and difficult to predict. To support this effort, we point developers to the additional monitoring features we implemented in OSS-Fuzz, as described in the following \cref{sec:recommendations-oss-fuzz}.

\section{Detecting Harness Degradation} 
\label{sec:recommendations-oss-fuzz}

As we have identified common causes of coverage drops, we now introduce an approach to detect harness degradation.
This requires practical metrics that allow maintainers to monitor harness health over time.
This aligns with the approach of Fuzz Introspector and OSS-Fuzz, which already provide certain metrics to identify problems in the fuzzing process.
The most relevant existing metrics are:
\begin{inparaenum}[(1)]
    \item The project and the fuzzing harness should build and run successfully. Failures are reported as build errors in the issue tracker for OSS-Fuzz. Note that this does not include partial build failures.
    \item Bugs resulting from found crashes are reported to the issue tracker. As crashes can severely impede fuzzer effectiveness, they can be seen as a type of notification.
    \item Introspector reports a ``blocked fuzzer'' if the achieved code coverage is too low, under 30\%, compared to what is possible based on static analysis.
    \item Introspector also reports functions as blockers if they lead to unexplored code with large complexitythat the fuzzer cannot get through.
    This is less of a metric and more of guidance for maintainers on what to focus on to improve fuzzing.
\end{inparaenum}

These metrics are valuable for project maintainers, but our case study of coverage drops (see~\Cref{sec:rq-factors-harness-degradation}) revealed several blind spots. To address them, we collaborated with the developers of OSS-Fuzz and Fuzz Introspector to introduce additional metrics:
\begin{inparaenum}[(1)]
    \item Previously, timeouts, out-of-memory conditions, and similar errors during the coverage gathering process were not reported but ignored. Based on maintainer feedback, this behavior was unexpected. We implemented error detection, collection of error messages, and surfacing of this information in the Fuzz Introspector project overview~\cite{fuzzIntrospectorProjectsOverview}.
    \item Similarly, the statefulness of harnesses is not detected, which can be another cause of inconsistent coverage results.
    Note that this has been a planned feature of OSS-Fuzz before. %
    At the time of writing, we provided a pull request to OSS-Fuzz that implements an approach to detect statefulness, based on collecting coverage twice with a different order of execution for the corpus entries.
    However, at the time of writing, this pull request has not yet been merged, and we leave it to the developers of OSS-Fuzz to decide its fate.
    \item A closely related metric is the corpus size in relation to covered code or covered complexity.
    While coverage is a useful metric, it does not allow for assessing how well the covered code is fuzzed.
    For example, a single seed corpus entry may cover 80\% of the code, but there could be a data integrity check stopping the fuzzer from generating any new related inputs.
    So, while the coverage percentage is at an acceptable level, based on the seed corpus alone, the project is effectively not being fuzzed.
    To mitigate this, we implemented the reporting of corpus size in Fuzz Introspector. A more advanced metric, such as relating the number of corpus entries to covered complexity or lines, remains an interesting direction for future work.
    A study of this relationship might also prove to be interesting, which, however, exceeds the scope of our current study.
\end{inparaenum}

Note that neither existing nor newly introduced metrics capture the temporal aspect, as they do not account for how harness effectiveness changes as a project evolves. To close this gap, we propose and implement additional metrics specifically designed to detect harness degradation over time:
\begin{inparaenum}
    \item[(4)] We keep track of past project fuzz targets and inform maintainers if any are no longer available.
    While OSS-Fuzz warns of full project build failures, it misses cases where only individual fuzz targets stop building.
    To address this, we implemented a mechanism that keeps track of all past fuzz targets and issues a warning in the Fuzz Introspector project overview if a target disappears. Previously, such targets would simply vanish without notice. A possible extension would allow maintainers to specify an explicit list of expected harnesses, making it clear which fuzz targets must remain available.
    \item[(5)] We found that most harnesses produce remarkably stable coverage results, usually identical line coverage when the underlying code remains unchanged. This observation motivated a simple but effective metric: tracking coverage trends over time to flag anomalies. Instability in coverage usually indicates a broken measurement setup, stateful harness behavior, or genuine code changes, which are all important signals for maintainers. To support this, we extended the Fuzz Introspector project overview with warnings whenever coverage drops by more than 5\% compared to the maximum of the past 30 days, and with annotations for individual days showing such drops.
    Finally, to make this coverage-based metric meaningful, it is best to classify code as \emph{intended to be fuzzed} or not.
    However, this needs to be a decision by the maintainers.
    Still, a future improvement could be to view the core project code and each library code separately, but this will require further effort to accurately and robustly perform this separation of code coverage information. Additionally, to get truly representative coverage percentage values, dead-code elimination should be turned off, which, unfortunately, is not possible for the OSS-Fuzz data in retrospect. Alternatively, developers can utilize additional absolute coverage numbers to detect and circumvent pitfalls created by dead-code elimination, as we have done in \Cref{sec:rq-factors-harness-degradation}. %
    
\end{inparaenum}

\begin{table}[htb]
\centering
\caption{Fuzz Introspector --- Degradation Alerts}
\label{tab:degradation-alerts}
\begin{tabular}{lcccc}
\toprule
 & \textbf{Total} & \textbf{With Alert} & \textbf{Coverage: <30\% / >30\%} & \textbf{Over 30\% and Alert} \\
\midrule
Projects & 179 & 38 & -- & -- \\
Fuzz Targets & 1957 & 105 & 1665 / 292 & 18 \\
\bottomrule
\end{tabular}
\end{table}

\begin{wraptable}{r}{0.45\linewidth}
\centering
\caption{Issue Type Counts}
\label{tab:issue-types}
\begin{tabular}{lr}
\toprule
\textbf{Issue Type} & \textbf{Count} \\
\midrule
Coverage Errors (Combined) & 62 \\
Fuzz Target Unavailable & 37 \\
Coverage Degraded & 13 \\
\bottomrule
\end{tabular}
\end{wraptable}

At the time of writing, all our contributions, except for statefulness detection, have been added to OSS-Fuzz or Fuzz Introspector.
A potential future extension would be an active reporting system that notifies maintainers directly, rather than requiring them to regularly check their project overview pages. 
Whether to adopt such a system remains up to the OSS-Fuzz and Fuzz Introspector developers.

\Cref{tab:degradation-alerts} shows summary statistics of our implementation efforts, as of the end of July 2025.
The table reports the number of projects and how many of them have an alert. Note that only projects that can be built with the Fuzz Introspector instrumentation are analyzed. Additionally, the number of fuzz targets and their number of alerts is shown. We also show if the fuzz target is ``blocked'', i.e., it is below 30\% coverage, and those targets that are not blocked but only alerted because of our contributions. Note that with the added alerts, maintainers will also have more information than just being notified if the fuzz target is blocked.

Furthermore, we show the number of alert types in \Cref{tab:issue-types}.
We can see that coverage errors are leading, followed by possible build failures and coverage degradation errors.

\section{Related Work}

\paragraph{Effectiveness of fuzzing}
\citeauthor{StudyOSSFuzzBugs}~\cite{StudyOSSFuzzBugs} analyze the life cycle of bugs in OSS-Fuzz and identify spikes of rapid bug discovery (punctuated equilibria) amidst long periods of low bug-finding activity.
\citeauthor{zhu2021regressionGreyboxFuzzing}~\cite{zhu2021regressionGreyboxFuzzing} also studied the lifecycle of bugs discovered by OSS-Fuzz and found that after an initial burst of bugs found as soon as projects are integrated into OSS-Fuzz, new bugs continue to be found \emph{at a constant rate} throughout the lifetime of the project. Then, about 77\% of new bugs discovered are regressions (bugs introduced with a recent commit). The rate of regression bugs increases after the initial burst of found bugs. Other studies~\cite{bohme2020fuzzingExponentialCosts, klooster2022effectivenessFuzzingCiCdPipelines} demonstrate diminishing returns even within a fuzzing campaign as it continues for a long time. In contrast, we examine whether harnesses are still maintained within OSS-Fuzz and whether fuzzer effectiveness may degrade over time. \citeauthor{nourrymyFuzzingBuildFails}~\cite{nourrymyFuzzingBuildFails} analyze build failures in OSS-Fuzz and found that only 5\% of builds for the median project fail, and that 80\% of those failures are fixed within one day, which shows that fuzzing mostly remains functional over a project's lifetime. As reasons for build failures, they identify environment, project dependency, and configuration issues, amongst others. \citeauthor{nourry2023human}~\cite{nourry2023human} studied Github Issues to understand which challenges developers face. They found, for instance, that developers find writing good harnesses hard, or that projects might suddenly fail to build for project-unspecific reasons, like changes to the fuzzing framework.

\paragraph{Automatic harnessing}
Writing good fuzzing harnesses is tedious and remains a key challenge in fuzzer integration~\cite{liang2018fuzzChallengesPractice, nourry2023human}, motivating efforts to automate their generation~\cite{bohme2020fuzzingChallengesReflections, yan2022surveyHumanMachineCollabFuzzing}. This resulted in various attempts to automate the process, including static analysis~\cite{zhang2021intelligen, zhang2023automataGuidedGen, ossFuzzGenJavaPaper}, guidance by runtime data~\cite{jung2021winnieHarnessSynthesis, zhang2023daisyFuzzDriverGenDynamic, jeon2023spHarnessSynthesis}, or a combination of both dynamic and static approaches~\cite{zhang2021apicraft}. Others suggest using artifacts such as unit tests~\cite{jeong2023utopia} or client code for libraries~\cite{babic2019fudgeHarnessSynthesis, ispoglou2020fuzzgenHarnessSynthesis}. Finally, LLMs have been proposed~\cite{lyu2023promptFuzzingForDriverGen} to generate fuzz harnesses. This includes efforts by OSS-Fuzz, which recently adapted techniques to generate fuzzing harnesses via LLMs~\cite{ossFuzzGen}. In contrast, our study motivates automatic harness \emph{maintenance} to address degradation and build failures over time.

\paragraph{Test suite degradation}
The degradation over time of unmaintained software components \cite{softwareDegradation1, softwareDegradation2, softwareDegradation3, softwareDegradation4, softwareDegradation5} or of unmaintained test suites ~\cite{alegroth2016maintenance, berner2005observations, karhu2009empiricalSoftwareTestingAutomation} has been well-studied in software engineering. Test suites do not only degrade like other software components; they also degrade when individual test cases become obsolete. Fuzzing as a testing technique is, by design, less prone to this kind of degradation, as test cases can automatically be added or removed by a fuzzer if the program under test changes. Our findings further underline the robustness of fuzzing as a method of test case generation, specifically regarding the aspect of harness degradation.

\section{Conclusion}

In this paper, we presented the first systematic study of fuzz harness degradation in OSS-Fuzz. By analyzing coverage and bug-finding capability over time and investigating coverage drops, we identified common causes of degradation.
Our results confirm that fuzzing is highly effective in the long term~\cite{bohme2020fuzzingChallengesReflections, ossFuzzRepo}, with many harnesses showing a surprisingly long effective lifetime where bugs are still found. However, actively maintained harnesses achieve higher coverage and uncover more bugs, while silent degradation can erode effectiveness unnoticed. To counter this risk, we proposed and implemented new monitoring measures for OSS-Fuzz and Fuzz Introspector that detect harness degradation early. Furthermore, we provide concrete, actionable guidelines for developers and recommend to consider harness updates after the addition of new features, changes to the parsing or checking of program inputs, and major version jumps. Furthermore, we advise developers to carefully evaluate changes to harnesses with the help of relative and absolute coverage metrics.
The main takeaway is that fuzzing delivers lasting security benefits, but only if harnesses are maintained and monitored to avoid silent degradation.

\section*{Data Availability}

All artifacts (dataset, case study notes, scraping code, and analysis notebook) are publicly available at \datasetDOI~and \url{\repoURL}.

\begin{acks}
We thank Marcel Böhme for insightful discussions that shaped this work, and Moritz Bley, Bhupendra Acharya, Maximilian Golla, and Avian Krämer for feedback and proofreading. We also thank the anonymous reviewers for their constructive comments.
This work was supported by the
\grantsponsor{erc}{European Research Council}{https://doi.org/10.13039/100010663}
    under Grant No.~\grantnum{erc}{101045669} (ERC consolidator grant RS$^3$).
\end{acks}

\appendix

\bibliographystyle{ACM-Reference-Format}
\bibliography{strings,bibliography}

\renewcommand{\figurename}{\bf Fig.}
\renewcommand{\tablename}{\bf Table}

\end{document}